\documentclass[twocolumn,aps,prb,superscriptaddress]{revtex4-2}

\usepackage{amsfonts}
\usepackage{amsmath}
\usepackage{amssymb}
\usepackage[pdftex]{graphicx}
\usepackage{dcolumn}
\usepackage{bm}
\usepackage{braket}
\usepackage{color}
\usepackage{here}
\usepackage{mathrsfs}
\usepackage{mathtools}
\usepackage[normalem]{ulem}

\begin{document}

\title{Heat rectification through a quantum two-level system}

\author{Tsuyoshi Yamamoto}
\email{tsuyoshi.yamamoto@cea.fr}
\affiliation{Univ. Grenoble Alpes, CEA, Grenoble INP, IRIG, PHELIQS, 38000 Grenoble, France}
\affiliation{Institute for Solid State Physics, the University of Tokyo, Kashiwa, Chiba 277-8581, Japan}
\author{Manuel Houzet}
\affiliation{Univ. Grenoble Alpes, CEA, Grenoble INP, IRIG, PHELIQS, 38000 Grenoble, France}

\date{\today}

\begin{abstract}
We study heat rectification through a quantum two-level system asymmetrically coupled to two thermal baths, as described by the Ohmic spin-boson model.
We evaluate the steady-state heat current using a tensor-network approach, which enables us to access the strongly correlated regime, and benchmark the results against analytical formulas in several limiting regimes, including the weak-coupling and incoherent-tunneling regimes.
We identify a scaling regime where the studied system flows from an ultraviolet regime, at temperatures larger than the Kondo temperature, to an infrared regime, at temperatures lower than the Kondo temperature.
By applying perturbation theory near the infrared fixed point, we find that the rectification ratio follows a universal power law.
Our numerical results agree well with this analytical prediction.
Our results provide a fundamental understanding of how dissipation-induced many-body physics affects heat transport.
\end{abstract}

\pacs{Valid PACS appear here}

\maketitle

\section{Introduction}

Quantum heat transport in nanostructures has attracted broad interest as a fundamental problem in nonequilibrium quantum physics~\cite{Pekola2015, Pekola2021}.
A paradigmatic example is the quantization of thermal conductance~\cite{Pendry1983}, which has been experimentally observed in a variety of mesoscopic systems, including phonons in a nanobridge~\cite{Schwab2000}, electrons across a semiconducting quantum point contact~\cite{Jezouin2013}, and microwave photons in a superconducting circuit~\cite{Meschke2006, Timofeev2009}.
Understanding quantum heat transport is also essential for developing quantum thermal technologies, in which quantum effects in nanoscale systems play a central role~\cite{Pekola2021}.

Over the past two decades, superconducting circuits have provided a versatile platform for studying quantum heat transport, thanks to their high tunability by external control fields and the availability of precise thermometry at sub-kelvin temperatures~\cite{Giazotto2006, Giazotto2012, Ronzani2018, Maillet2020, Senior2020, Gubaydullin2022, Upadhyay2025}.
Moreover, superconducting circuits enable nonlinear elements to be coupled to high-impedance electromagnetic environments, thereby providing access to nontrivial many-body phenomena induced by dissipation, such as a bosonic version of the Kondo physics~\cite{LeHur2012, Goldstein2013, Saito2013} and various quantum phase transitions~\cite{Yamamoto2018PRB, Filippis2023, Schmid1983, Houzet2024, Burstein2024, Yamamoto2024, Paris2025}.
Recent experiments have started to measure heat transport through a nonlinear element strongly coupled to thermal baths~\cite{Subero2023}, opening the door to experimental observation of dissipation-induced many-body phenomena in superconducting-circuit platforms.

A quantum two-level system is the simplest nonlinear quantum system and constitutes a basic building block of superconducting circuits.
Superconducting qubits realize effective two-level systems, and several types have been proposed and demonstrated~\cite{Nakamura1999, Orlando1999, Martinis2002, Koch2007, Manucharyan2009}.
For example, the transmon qubit consists of a Josephson junction shunted by a large capacitance, which suppresses charge noise while keeping anharmonicity~\cite{Koch2007}.
When such a two-level system is coupled to photonic thermal baths, such as transmission lines or normal-metal reservoirs, the resulting setup can be modeled by the Ohmic spin-boson model~\cite{Leggett1987, Weiss_text}.
This model is relevant not only to superconducting qubits~\cite{Magazzu2018, Leppakangas2018} but also to anharmonic molecular junctions~\cite{Nitzan_text} and cold atoms~\cite{Recati2005}.
Despite its apparent simplicity, heat transport in the Ohmic spin-boson model exhibits a remarkably rich range of behavior depending on system-bath coupling strength and temperature.
Near thermal equilibrium, steady-state heat transport in the Ohmic spin-boson model has been systematically studied using quantum Monte Carlo simulations~\cite{Saito2013, Yamamoto2018}.
By contrast, steady-state heat transport far from equilibrium remains less understood.

\begin{figure}
    \centering
    \includegraphics[width=1.0\linewidth]{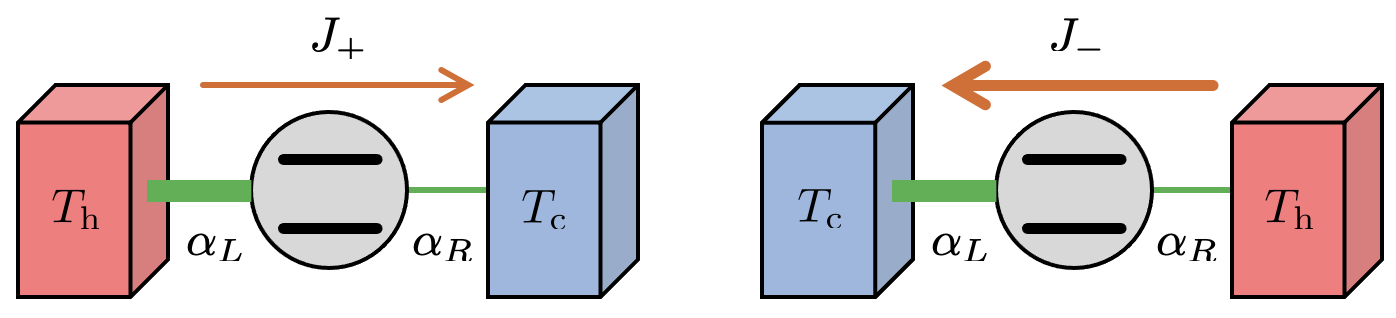}
    \caption{Quantum heat transport through a two-level system coupled to two thermal baths at different temperatures. When the left bath is hot ($T_{\rm h}$) and the right bath is cold ($T_{\rm c}$), the heat current $J_+$ flows from left to right (forward direction; left panel). Upon exchanging the two bath temperatures, the heat current $J_-$ flows in the opposite direction (backward direction; right panel). Asymmetric system-bath coupling ($\alpha_L \ne \alpha_R$) can lead to heat rectification, unequal magnitudes of the forward and backward heat currents, $|J_+|\ne |J_-|$. For a fixed coupling asymmetry, e.g., $\alpha_L>\alpha_R$, we identify regimes where the rectification ratio, $\mathcal{R}=|J_+|/|J_-|$, can be smaller or larger than unity.}
    \label{fig:setup}
\end{figure}

A prominent hallmark of nonequilibrium heat transport beyond the linear response is heat rectification.
In a setup with two thermal baths, heat rectification means that, upon exchanging the temperatures of the two thermal baths, the heat current is not only reversed, but also its magnitude changes (see Fig.~\ref{fig:setup}).
Its emergence requires three ingredients: (i) nonlinearity of the quantum link, (ii) spatial asymmetry in the system-bath couplings, and (iii) operation beyond the linear-response regime.
Heat rectification through a two-level system has been studied theoretically in several limiting regimes, including weak system-bath coupling~\cite{Velizhanin2010, Ruokola2011} and incoherent transport~\cite{Nicolin2011, Segal2014, Wang2015}.
By contrast, the strongly correlated regime, where dissipative many-body effects are expected to be most pronounced, remains comparatively underexplored.

In this work, we investigate heat rectification in the Ohmic spin-boson model over a broad range of dissipation strengths, with particular emphasis on the strongly-interacting regime.
To this end, we employ the tensor-network approach, the time-evolving matrix product operator (TEMPO) algorithm~\cite{Strathearn2018}, based on the Feynman-Vernon path-integral formalism~\cite{Feynman1963}.
By representing the real-time propagator dressed by the influence of the baths as a matrix product operator, TEMPO enables efficient computation of the reduced density matrix and the corresponding correlation function even under nonequilibrium conditions~\cite{Popovic2021, Dunnett2021, Chen_2023}.
The correlation function of the two-level system enables us to evaluate the steady-state heat current.
We show that the rectification ratio, defined as the ratio of the heat currents obtained upon exchanging the temperatures of the two thermal baths, exhibits a nontrivial dependence on the system-bath coupling at low temperatures.
Furthermore, by comparing the numerical results with analytical expressions derived from the effective Hamiltonian near the infrared (IR) fixed point, as well as with results obtained in other analytical regimes, we demonstrate that heat rectification persists even in the extremely low-temperature regime, beyond the reach of weak-coupling perturbative approaches.

This paper is organized as follows. In Sec.~\ref{sec:model}, we briefly review the equivalence between the Ohmic spin-boson model and the anisotropic Kondo model, and derive an exact expression for the heat current.
In Sec.~\ref{sec:analytics}, we present analytical results for the heat current and the linear thermal conductance in several regimes: the weak system-bath coupling regime, the incoherent-tunneling regime, and the perturbative regime near the IR fixed point.
In Sec.~\ref{sec:numerics}, we present the tensor-network results, benchmark them against the linear thermal conductance, and then discuss our main results for heat rectification.
Section~\ref{sec:summary} is devoted to the summary and discussion.

\section{Model}
\label{sec:model}

In this section, we introduce the Ohmic spin-boson model as a description of a two-level system coupled to two bosonic thermal baths, and then recall its relation with the anisotropic Kondo Hamiltonian.
We formulate the steady-state heat current and the linear thermal conductance, and define the rectification ratio that characterizes heat rectification.

\subsection{Ohmic spin-boson Hamiltonian}

A two-level system coupled to two bosonic baths is described by the spin-boson model (see Fig.~\ref{fig:setup})~\cite{Leggett1987, Weiss_text}
\begin{align}
\label{eq:Hsb}
H&=\frac{\Delta}{2}\sigma_x+\sum_{r=L,R}H_{B,r}+\sum_{r=L,R}H_{I,r}, \\
H_{B,r}&=\sum_{k}\omega_{r,k}b_{r,k}^\dagger b_{r,k}, \\
H_{I,r}&=\frac{\sigma_z}{2}\sum_{k}\lambda_{rk}(b_{r,k}+b_{r,k}^\dagger).
\end{align}
The first term in Eq.~\eqref{eq:Hsb} describes the hybridization of two levels with tunneling amplitude $\Delta$ and Pauli matrices $\sigma_{x,y,z}$ acting in the Hilbert space of the two-level system.
The left ($L$) and right ($R$) bosonic thermal baths are modeled as a collection of harmonic oscillators with frequency $\omega_{r,k}$ and annihilation operator $b_{r,k}$ of mode $k$ in bath $r=L,R$.
The interaction strength between the two-level system and the bath $r$ is $\lambda_{r,k}$.
The properties of the baths are characterized by the Ohmic spectral density,
\begin{align}
I_r(\omega)
&=\sum_k\lambda_{r,k}^2\delta(\omega-\omega_{r,k})
=2\alpha_rI_0(\omega), \\
I_0(\omega)
&=\omega e^{-\omega/\omega_c},
\end{align}
for $\omega>0$, and $I_r(\omega)=0$ otherwise, where $\alpha_r$ is the dimensionless coupling strength and $\omega_c$ is the cutoff frequency.

\subsection{Anisotropic Kondo Hamiltonian}

A unitary transformation that defines two linear superpositions ``$\pm$'' of the bosonic operators of the thermal baths,
\begin{align}
b_{\pm,k}
=\frac{\sqrt{\alpha_L}b_{L/R,k}\pm\sqrt{\alpha_R}b_{R/L,k}}{\sqrt{\alpha}},
\end{align}
where $\alpha=\alpha_L+\alpha_R$ is the total coupling strength, then allows decoupling one of these superpositions from the two-level system, $H=H_++H_-$ with
\begin{subequations}
\begin{align}
\label{eq:H+}
H_+&=\frac{\Delta}{2}\sigma_x+\sum_{k}\omega_kb_{+,k}^\dagger b_{+,k} \nonumber\\
&\qquad+\frac{\sigma_z}{2}\sum_{k}\lambda_{+,k}(b_{+,k}+b_{+,k}^\dagger), \\
H_-&=\sum_{k}\omega_kb_{-,k}^\dagger b_{-,k},
\end{align}
\end{subequations}
where $\lambda_{+,k}=\sqrt{\alpha/\alpha_L}~\lambda_{L,k}$, resulting in the Ohmic spectral density and $I_+(\omega)=\sum_k\lambda_{+,k}^2\delta(\omega-\omega_k)=2\alpha I_0(\omega)$; here we assume that the bosonic modes have a common linear dispersion $\omega_k=vk$ with the plasma velocity $v$.

Now, after the convenient phase redefinition $b_{\pm,k}\to e^{i\pi/2}b_{\pm,k}$, we introduce continuous bosonic fields,
\begin{subequations}
\begin{align}
\phi_\pm(x)
&=\frac{1}{\sqrt{L}}\sum_{n>0}\sqrt{\frac{\pi}{k_n}}\cos(k_nx)(b_{\pm,k}+b_{\pm,k}^\dagger), \\
\rho_\pm(x)
&=\frac{1}{i}\frac{1}{\sqrt{L}}\sum_{n>0}\sqrt{\frac{k_n}{\pi}}\cos(k_nx)(b_{\pm,k}-b_{\pm,k}^\dagger),
\end{align}    
\end{subequations}
where $k_n=n\pi/L$ ($n=1,2,\dots$) is the wave number and $L$ is a normalization length of the thermal baths.
These two fields satisfy the no-current boundary condition ($\partial_x\phi_\pm(x=0)=\partial_x\rho_\pm(x=0)=0$) and the commutation relation, $[\phi_s(x),\rho_{s'}(x')]=i\delta_{s,s'}\delta(x-x')$.
In the low-energy regime $\omega \ll \omega_c$ and in the limit $L\to\infty$, the Ohmic spin-boson model can be described by the low-energy effective Hamiltonian
\begin{align}
H
=H_-^0+H_+^0+\frac{\Delta}{2}\sigma_x-\frac{\pi v}{2}\sqrt{2\alpha}\rho_+(0)\sigma_z,
\end{align}
where $H_\pm^0$ is the free bosonic Hamiltonian,
\begin{align}
H_\pm^0
=\frac{v}{2\pi}\int_0^\infty dx~\Big\{[\partial_x\phi_\pm(x)]^2+[\pi\rho_\pm(x)]^2\Big\}.
\end{align}
Furthermore, by applying the unitary operator $U_+=\exp[-i\sigma_z\sqrt{\alpha/2}~\phi_+(0)]$, we obtain the bosonized form of the anisotropic Kondo Hamiltonian~\cite{Leggett1987, Weiss_text}
\begin{align}
\label{eq:HAK}
H'=H_-^0+H_+^0+\frac{\Delta}{2}\left[\sigma_+e^{+i\sqrt{2\alpha}\phi_+(0)}+{\rm h.c.}\right],
\end{align}
where $\sigma_\pm=(\sigma_x\pm i\sigma_y)/2$ are the spin ladder operators.
Note that the transformed Hamiltonian~\eqref{eq:HAK} is the sum of a free Hamiltonian and an Ohmic spin-boson model with an effective bath characterized by dimensionless coupling strength $\alpha$.
The scaling dimension of the interaction between the two-level system and the ``+'' bosonic modes is $\alpha$.
In the standard fermionic anisotropic Kondo model, a local impurity spin is coupled to the spin of the conduction electrons through anisotropic exchange interactions, with different coupling strengths in the longitudinal and transverse directions.
The correspondence between the anisotropic Kondo model and the Ohmic spin-boson model identifies the longitudinal and transverse Kondo couplings as $J_\parallel\propto\tan[\pi(1-\sqrt{\alpha})/2]$ and $J_{\perp}\propto\Delta/\omega_c$, respectively~\cite {Guinea1985, Costi1998}.
For $\alpha>1$, the longitudinal Kondo coupling is ferromagnetic ($J_\parallel<0$).
In the ferromagnetic region, when the transverse Kondo coupling $J_{\perp}$ lies below the critical line, the coupling is irrelevant and flows to the line of the ultraviolet (UV) fixed points ($J_\perp=0$, $J_\parallel<0$), where the two-level system is effectively decoupled from the thermal baths.
By contrast, when the transverse coupling exceeds the critical value, the couplings flow into the IR fixed point ($J_\perp\to\infty$, $J_\parallel\to\infty$), where the two-level system locks into a singlet state by strongly hybridizing with the thermal baths.
For $\alpha<1$, the longitudinal coupling is antiferromagnetic ($J_\parallel>0$).
In the antiferromagnetic region, the Kondo temperature $T_K$ is a crossover temperature that separates the high-temperature and low-temperature regimes governed by the UV and IR fixed points, respectively.
Note that, in the limit $\Delta/\omega_c\to0$ ($J_\perp\to0$), which we tacitely assumed, the Ohmic spin-boson model displays the phase transition between the singlet and doublet ground states at $\alpha=1$~\cite{Leggett1987, Weiss_text}.
The Kondo temperature at $\alpha<1$ is~\cite{Lesage1999npb, Freton2013} 
\begin{align}
\label{eq:TK}
T_K=\frac{\alpha}{2\sqrt{\pi}}\frac{\Gamma(\frac{\alpha}{2(1-\alpha)})}{\Gamma(\frac{1}{2(1-\alpha)})}\left[\Gamma(1-\alpha)\right]^{1/(1-\alpha)}\Delta\left(\frac{\Delta}{\omega_c}\right)^{\alpha/(1-\alpha)},
\end{align}
where $\Gamma(z)$ is the gamma function.
Note that the cutoff frequency $\omega_c$ enters the definition of the Kondo temperature~\eqref{eq:TK} up to a numerical prefactor, which depends on details of the cutoff.
Some limiting values of the Kondo temperature are $T_K(\alpha\to0)=\Delta/\pi$ and $T_K(\alpha=1/2)=\pi\Delta^2/(4\omega_c)$, while $T_K(\alpha)$ vanishes exponentially as $\alpha\to1$.
The Kondo temperature can also be related to the zero-temperature static susceptibility, $\chi_z=\partial_\epsilon\braket{\sigma_z}|_{\epsilon=0,T=0}$, through $T_K=2/(\pi\chi_z)$~\cite{Hewson_text, Lesage1999prl}, where $\epsilon$ is the detuning energy that would enter the Hamiltonian~\eqref{eq:H+} by adding the detuning term $-\epsilon\sigma_z$.

\subsection{Heat current}

The heat current flowing out of thermal bath $r$ is defined by
\begin{align}
J_r\equiv-\Braket{\frac{dH_{B,r}}{dt}}
={\rm Im}\sum_{k}\lambda_{r,k}\omega_{r,k}\braket{\sigma_z b_{r,k}}.
\end{align}
where $\braket{\cdot}$ denotes the quantum mechanical average about the non-equilibrium steady state of the total system.  
The Keldysh formalism provides the Meir-Wingreen formula~\cite{Meir1992} for the steady-state heat current $J\equiv J_L=-J_R$~\cite{Ojanen2008, Saito2013, Yamamoto2018},
\begin{align}
\label{eq:J}
J=\frac{\alpha_L\alpha_R}{\alpha}\int_0^\infty d\omega~\omega I_0(\omega)\chi''(\omega)[n_L(\omega)-n_R(\omega)],
\end{align}
where $n_r(\omega)=1/(e^{\omega/T_r}-1)$ is the Bose-Einstein distribution function of the bath $r$ with the temperature $T_r$.
Here, $\chi''(\omega)$ is the imaginary part of the dynamic susceptibility; the Fourier transformation of the correlation function $\chi(t-t')=i\theta(t-t')\braket{[\sigma_z(t),\sigma_z(t')]}$.
Its real part in the zero-frequency limit is related to the static susceptibility as $\chi'(0)=\chi_z$. 
When the bath temperatures are different ($T_L\neq T_R$), this correlation function must be evaluated under nonequilibrium conditions.

\subsection{Linear thermal conductance}

For the small temperature difference, $\delta T\ll T$, with $\delta T=T_L-T_R$ and $T=(T_L+T_R)/2$, the heat current is expanded in a series of the temperature difference as $J\approx G_{\rm th}\delta T+\mathcal{O}(\delta T)^2$.
The linear thermal conductance, $G_{\rm th}\equiv\lim_{\delta T\to0}J/\delta T$, is related with the dynamic susceptibility evaluated under thermal equilibrium at temperature $T$,
\begin{align}
\label{eq:Gth}
G_{\rm th}
=\frac{\alpha_L\alpha_R}{\alpha}\int_0^\infty d\omega~I_0(\omega)\chi''(\omega)\left[\frac{\omega/(2T)}{\sinh(\omega/2T)}\right]^2.
\end{align}
Note that the thermal conductance quantum multiplied by the coupling asymmetry factor, $G_{{\rm th},0}=4\alpha_L\alpha_R/\alpha^2\times \pi T/6$, would be realized at $T\ll\omega_c$ if $\chi''(\omega)=2/(\pi \alpha \omega)$.
We will see however that the thermal conductance always remains below this bound for the studied model

\subsection{Rectification ratio}

To discuss heat rectification, we consider the rectification ratio of the steady-state heat current in the forward ($J_+$) and backward ($J_-$) directions,
\begin{align}
\label{eq:R_def}
\mathcal{R}\equiv\frac{|J_+|}{|J_-|}=\frac{|J(T_L=T_{\rm h},T_R=T_{\rm c})|}{|J(T_L=T_{\rm c},T_R=T_{\rm h})|},
\end{align}
where $T_{\rm h}=T+|\delta T|/2$ and $T_{\rm c}=T-|\delta T|/2$ are hot and cold temperatures.
A non-unity rectification ratio $\mathcal{R}\ne1$ corresponds to heat rectification.
Without loss of generality, we take the coupling to the left thermal bath to be stronger than that to the right one, ($\alpha_L\ge\alpha_R$).
With this convention, $\mathcal{R}>1$ indicates that a larger heat current flows when the two-level system is more strongly coupled to the hot bath than to the cold bath.
Conversely, $\mathcal{R}<1$ indicates that a larger heat current flows when the two-level system is more strongly coupled to the cold bath than to the hot bath.

\section{Analytical formula}
\label{sec:analytics}

In this section, we first review analytical formulas for the steady-state heat current, the linear thermal conductance, and the rectification ratio in several regimes: the weak system-bath coupling regime, the incoherent-tunneling regime, and the Toulouse point.
We then derive their low-temperature asymptotic behavior in the perturbative regime near the IR fixed point.

\subsection{Weak-coupling regime ($\alpha\ll1$)}
\label{sec:weak}

For the weak coupling between the two-level system and the baths ($\alpha_r\ll1$), we can treat the interaction Hamiltonian $H_{I,r}$ as a perturbation.
In this regime, heat is transferred through an energy-conserving process in which a boson is absorbed from one thermal bath and emitted into the other.
We here evaluate the steady-state heat current from the left to right thermal baths, representing this process, rather than the general expression~\eqref{eq:J}, as
\begin{align}
J_{L\to R}=P_0\sum_{i,f}\omega_{L,k_i}\Gamma_{i\to f}W_i,
\end{align}
where $P_0$ is the probability that the two-level system is in its ground state, $\omega_{L,k_i}$ is the energy of the boson absorbed from the left thermal bath,  $\Gamma_{i\to f}$ is the transition rate from the initial state to the final state, and $W_{i}\propto e^{-\omega_{L,k_i}/T_L}e^{-\omega_{R,k_i}/T_R}$ is the thermal distribution of the initial state.
Using the generalized Fermi's golden rule to the interaction Hamiltonian to evaluate the transition rate, the net heat current is then given by $J=J_{L\to R}-J_{R\to L}$~\cite{Ruokola2011},
\begin{align}
\label{eq:Jwc}
J&=\frac{\pi\alpha_L\alpha_R}{2}P_0\int_0^\infty d\omega~\omega^3[n_L(\omega)-n_R(\omega)] \nonumber\\
&\qquad\qquad
\times\left|\frac{1}{\omega-\Delta+i\Gamma_1/2}-\frac{1}{\omega+\Delta-i\Gamma_1/2}\right|^2,
\end{align}
with $P_0=\Gamma_1/(\Gamma_0+\Gamma_1)$, where $\Gamma_0=\pi I_0(\Delta)\sum_{r}\alpha_r n_r(\Delta)$ and
$\Gamma_1=\pi I_0(\Delta)\sum_r\alpha_r [n_r(\Delta)+1]$
denote the excitation rate of the two-level system from the ground state due to the thermal baths and the decay rate of the excited two-level system into the thermal baths, respectively.

At high temperatures ($T_r>\Delta$), heat transport is dominated by sequential tunneling, in which thermal bosons
with energy $\Delta$ from the thermal baths are sequentially absorbed by and emitted from the two-level system.
In the high-temperature limit ($T_r\gg\Delta$), since the resonant energy transfer at $\omega=\Delta$ becomes dominant, and thus we can approximate the last factor $|\omega\mp\Delta\pm i\Gamma_1/2|^{-2} $ as $(2\pi/\Gamma_1)\delta(\omega\mp\Delta)$ in Eq.~\eqref{eq:Jwc}, the steady-state heat current is obtained as
\begin{align}
J\approx\frac{\pi^2T_K^2}{2}\frac{\alpha_L\alpha_R}{\alpha_LT_L+\alpha_RT_R}(T_L-T_R),
\end{align}
and the thermal conductance is
\begin{align}
G_{\rm th}\approx \frac{\pi^3T_K^2}{2}\frac{\alpha_L\alpha_R}{\alpha}\frac{1}{T}
\end{align}
Note that the Kondo temperature~\eqref{eq:TK} is reduced to $\Delta/\pi$ at the weak coupling limit $\alpha\to0$.
The expression for the steady-state heat current implies heat rectification, and then the rectification ratio is given by
\begin{align}
\label{eq:Rwc}
\mathcal{R}\approx1-\frac{2\delta\tilde{\alpha}|\delta\tilde{T}|}{2+\delta\tilde{\alpha} |\delta \tilde{T}|},
\end{align}
where $\delta\tilde{\alpha}=\delta\alpha/\alpha=(\alpha_L-\alpha_R)/\alpha$ and $\delta\tilde{T}=\delta T/T=(T_L-T_R)/T$.
This result shows that, for $\alpha_L>\alpha_R$, the rectification ratio is always smaller than unity, and approaches zero in the limits $\alpha_R\ll\alpha_L$ and $T_{\rm c}\ll T_{\rm h}$.

As the temperature decreases, sequential tunneling is exponentially suppressed because the thermal bosons no longer have sufficient energy to excite the two-level system.
At low temperatures ($T_r\ll\Delta$), two-boson cotunneling involving virtual excitations becomes dominant.
For this process, we can neglect the $\omega$ dependence and $\Gamma_1$ in the last factor in Eq.~\eqref{eq:Jwc}, and then the steady-state heat current is approximated as
\begin{align}
\label{eq:Jcotunn}
J\approx\frac{2}{15}\pi^3\frac{\alpha_L\alpha_R}{T_K^2}\left(T_L^4-T_R^4\right),
\end{align}
and the linear thermal conductance is
\begin{align}
\label{eq:Gwc}
G_{\rm th}\approx \frac{8}{15}\pi^3\frac{\alpha_L\alpha_R}{T_K^2}T^3.
\end{align}
The cubic behavior in the temperature reflects the linear suppression of the correlation function at low frequencies, $\chi''(\omega\ll\Delta)\propto \omega$.
As a result, the thermal conductance is strongly reduced with respect to the thermal conductance quantum, $G_{\rm th}/G_{{\rm th},0}=(4/5)\pi^2\alpha^2(T/T_K)^2$.
Since the steady-state heat current is symmetric with respect to the exchange of the left and right couplings (see Eq.~\eqref{eq:Jcotunn}), it does not produce heat rectification. Therefore, the rectification ratio approaches unity exponentially in the low-temperature limit, $T_r\ll\Delta$.

\subsection{Incoherent-tunneling regime ($\alpha<1$ and $T>T_K$ or $\alpha>1$)}
\label{sec:niba}

In Sec.~\ref{sec:weak}, we obtained the analytical formula for the weak coupling regime.
Now, let us consider the opposite regime, strong coupling and/or high temperature.
In the high-temperature regime, bath-induced thermal fluctuations dominate over coherent quantum tunneling.
In the strong-dissipation regime ($\alpha>1$), Ohmic dissipation suppresses coherent superpositions of the localized states.
Consequently, for $\alpha<1$ at temperatures above the Kondo scale, $T\gtrsim T_K$, and for $\alpha>1$, the dynamics is incoherent and can be described within the noninteracting-blip approximation (NIBA)~\cite{Leggett1987, Weiss_text}.
The NIBA is an approximation formulated within the real-time path-integral approach to the spin-boson model.
{In this formulation, the real-time evolution of the reduced density matrix of the two-level system is represented in terms of sojourns and blips, which denote time intervals corresponding to its diagonal and off-diagonal elements, respectively.
The NIBA neglects bath-induced interactions between different blips, assuming that tunneling events are sufficiently dilute, which allows one to write down a closed form of the generalized master equation for the dynamics.

Within the NIBA, the steady-state heat current is expressed as~\cite{Segal2014}
\begin{align}
\label{eq:Jniba}
J=\frac{\Delta^2}{8\pi}\int_0^\infty d\omega~\omega[k_R(\omega)k_L(-\omega)-k_R(-\omega)k_L(\omega)],
\end{align}
where $k_r(\omega)$ is the frequency-dependent transition rate,
\begin{align}
\label{eq:k}
k_r(\omega)=2{\rm Re}\int_0^\infty dt~e^{i\omega t}e^{-Q_r(t)},
\end{align}
and it satisfies $k_r(-\omega)=e^{-\omega/T_r}k_r(\omega)$.
Here, $Q_r(t)=Q'_r(t)+iQ''_r(t)$ is the complex correlation function,
\begin{subequations}
\label{eq:Q}
\begin{align}
Q'_r(t)
&=2\alpha_r\int_0^\infty d\omega~I_0(\omega)\frac{1-\cos(\omega t)}{\omega^2}\coth\left(\frac{\omega}{2T_r}\right), \\
Q''_r(t)
&=2\alpha_r\int_0^\infty d\omega~I_0(\omega)\frac{\sin(\omega t)}{\omega^2}.
\end{align}
\end{subequations}
Note that the NIBA is equivalent to the Redfield master equation applied to the polaron-transformed Hamiltonian, $H_{\rm P}=U_{\rm P}^\dagger HU_{\rm P}=\sum_{r}H_{B,r}-(\Delta/2)(\sigma_+e^{\eta}+\sigma_-e^{-\eta})$ with $U_{\rm P}=e^{-\sigma_z\eta/2}$ and $\eta=\sum_{rk}(\lambda_{rk}/\omega_{rk})(b_{rk}^\dagger-b_{rk})$, in which the transformed system-bath interaction, $\propto\Delta$, can be treated perturbatively~\cite{Wang2015}.

In the scaling regime ($T_K\ll\max(T,~\omega)\ll\omega_c$), the frequency-dependent transition rate~\eqref{eq:k} takes scaling form~\cite{Segal2014}
\begin{align}
k_r(\omega)
=\frac{1}{\omega_c}\left(\frac{\omega_c}{2\pi T_r}\right)^{1-2\alpha_r}\frac{|\Gamma(\alpha_r+i\omega/(2\pi T_r))|^2}{\Gamma(2\alpha_r)}e^{\omega/(2T_r)},
\end{align}
resulting that the thermal conductance shows the $\alpha$-dependent temperature dependence, $G_{\rm th}(T\ll \omega_c)\approx a(\alpha)\Delta^2/\omega_c(T/\omega_c)^{2\alpha-1}$ with $a(\alpha)=(2\pi)^{2\alpha+1}\alpha_L\alpha_R\Gamma^2(\alpha)/[8(2\alpha+1)\Gamma(2\alpha)]$.
Note that, at $\alpha=1$, the thermal conductance in the scaling regime is $G_{\rm th}(T\ll\omega_c)\approx4\alpha_L\alpha_R/\alpha^2\times(\pi^3/12)(\Delta/\omega_c)^2T$, which is parametrically smaller than $G_{{\rm th},0}$.
The factor $(T_r/\omega_c)^{2\alpha_r-1}|\Gamma(\alpha_r+i\omega/(2\pi T_r))|^2$ in the transition rate induces heat rectification.
Since the contribution from the gamma function is relatively weak, the steady-state heat current is governed mainly by the factor $(T_r/\omega_c)^{2\alpha_r-1}$.
Thus, the rectification ratio can be roughly estimated as $\mathcal{R}\sim(T_{\rm h}/T_{\rm c})^{2(\alpha_L-\alpha_R)}$.
It implies $\mathcal{R}>1$ for $\alpha_L>\alpha_R$ in the scaling regime $T_K\ll T_r\ll\omega_c$.
For $\alpha>1$, since the Kondo temperature vanishes, this temperature window is accessible.
In contrast, for $\alpha<1$, the finite Kondo temperature reduces the available scaling window, making this regime difficult to access.
In the limit $T_r\to0$ for $\alpha_r>1$, the frequency-dependent rate reduces to a temperature-independent form, $k_r(\omega)\sim(2\pi/\omega_c)(\omega/\omega_c)^{2\alpha_r-1}/\Gamma(2\alpha_r)$.
As a result, heat rectification vanishes in this limit, since the heat current becomes insensitive to swapping the bath temperatures.

For $\alpha<1$, the thermal conductance can be scaled by the Kondo temperature as $G_{\rm th}(T_K\ll T\ll\omega_c)\approx\tilde{a}(\alpha) T_K(T/T_K)^{2\alpha-1}$, where $\tilde{a}(\alpha)=a(\alpha)[\Gamma(1-\alpha)]^{-2}[\alpha/(2\sqrt{\pi})\Gamma(\alpha/2(1-\alpha))/\Gamma(1/2(1-\alpha))]^{2(\alpha-1)}$ is the $\alpha$-dependent prefactor.
In particular, at $\alpha=1/2$, the thermal conductance becomes independent of the temperature, $G_{\rm th}=\pi^2\alpha_L\alpha_RT_K$.

Finally, we note that we numerically perform the integrations in Eqs.~\eqref{eq:Jniba}, \eqref{eq:k}, and \eqref{eq:Q} to evaluate the thermal conductance and rectification ratio for comparison with the numerical results presented in Sec.~\ref{sec:numerics}.

\subsection{Toulouse point ($\alpha=1/2$)}

At $\alpha=1/2$, known as the Toulouse point, by mapping to the noninteracting resonant-level model, we can exactly solve the correlation function as~\cite{Weiss_text}
\begin{align}
\label{eq:chi_toulouse}
\chi(\omega)=\frac{4}{\pi\omega}\frac{\gamma}{\omega+i\gamma}\left[\psi(x)-\psi\left(x-i\frac{\omega}{2\pi T}\right)\right],
\end{align}
where $\psi(z)$ is the digamma function with $x=1/2+\gamma/(4\pi T)$ and $\gamma=\pi\Delta^2/(2\Omega_{\rm T})$.
Here, $\Omega_{\rm T}$ is the cutoff frequency of the noninteracting resonant-level model.
The characteristic frequency is compatible with the Kondo temperature scale, $\gamma=2T_K\times \omega_c/\Omega_{\rm T}$.
By substituting this thermal-equilibrium exact solution into the formula for the linear thermal conductance~\eqref{eq:Gth}, we obtain low- and high-temperature asymptotes as
\begin{subequations}
\begin{align}
G_{\rm th}(T\ll\gamma)
&\approx\frac{128}{15}\pi^3\alpha_L\alpha_R\gamma\left(\frac{T}{\gamma}\right)^3. \\
G_{\rm th}(T\gg\gamma)
&\approx\pi^2\alpha_L\alpha_R\frac{\gamma}{2},
\end{align}
\end{subequations}
respectively.
The temperature-independent thermal conductance in the high-temperature regime is consistent with the NIBA result at $\alpha=1/2$.
Note that the above expression is valid in the low-energy regime, $\omega,~T\ll\Omega_{\rm T}$~\cite{Leggett1987}.
We also note that this Toulouse-point mapping is formulated for thermal equilibrium and therefore does not directly provide a way to compute heat rectification, which is intrinsically a nonequilibrium effect. Thus, the Toulouse-point result should be regarded mainly as an exactly solvable benchmark for linear heat transport rather than heat rectification.

\subsection{Perturbation regime around the IR fixed point}

So far, we provided the analytical formulas for the weak-coupling and incoherent-tunneling regimes.
However, they cannot capture the many-body effects for the strongly correlated regime ($\alpha\lesssim1$ and $T\ll T_K$).
Here, to see the many-body effects, we start from the anisotropic Kondo Hamiltonian~\eqref {eq:HAK} and consider the perturbation around the IR fixed point.

At the IR fixed point, the two-level system is strongly correlated to the thermal baths, and we reintroduce free bosonic fields $\tilde{\phi}(x)$ and $\tilde{\rho}(x)$.
Since the fermionic field operators are expressed as $\psi_\uparrow(x)\sim e^{+i[\tilde{\phi}(x)+\tilde{\theta}(x)]/\sqrt{2\pi}}$ and $\psi_\downarrow(x)\sim e^{-i[\tilde{\phi}(x)-\tilde{\theta}(x)]/\sqrt{2\pi}}$, where $\tilde{\theta}(x)$ is the dual field defined by $\tilde{\rho}(x)=-\partial_x\tilde{\theta}(x)/\pi$~\cite{Guinea1985}, a phase shift by $\delta$ of the fermionic operators at the boundary, $\psi_{\uparrow,\downarrow}(0_+)=\psi_{\uparrow,\downarrow}(0_-)e^{2i\delta}$, can be accounted for by a constant shift of the dual field, $\tilde{\theta}(0_+)=\tilde{\theta}(0_-)+2\delta\sqrt{2\pi}$.
At the IR fixed point, the Kondo phase shift ($\delta=\pi/2$) occurs, whereas $\delta=0$ at the UV fixed point.
Therefore, at the IR fixed point, the boundary conditions for $\tilde{\phi}(x)$ and $\tilde{\rho}(x)\propto\partial_x\tilde{\theta}(x)$ remain unchanged compared to the no-current boundary conditions at the UV fixed point.

The anisotropic Kondo Hamiltonian is expanded around the IR Hamiltonian $H_{\rm IR}$ in a series of irrelevant boundary operators $O_{2i}$ as
\begin{align}
\label{eq:H+expand}
H_+
&=H_{\rm IR}+T_K\sum_{i=1} c_{2i}\frac{O_{2i}}{(T_K/v)^{2i}} \nonumber\\
&\approx H_{\rm IR}+c_2T_K\frac{O_2}{(T_K/v)^2}+c_4T_K\frac{O_4}{(T_K/v)^4}, \\
H_{\rm IR}
&=\frac{v}{2\pi}\int_0^\infty dx~\Big\{[\partial_x\tilde{\phi}(x)]^2+[\pi\tilde{\rho}(x)]^2\Big\},
\end{align}
where the irrelevant operators are~\cite{Lesage1999npb, Freton2013, note_T}
\begin{subequations}
\begin{align}
O_2
&=\pi:\tilde{\rho}^2(0):, \\
\label{eq:O4}
O_4
&=\pi^3:\tilde{\rho}^4(0):-\frac{\pi}{2}:\tilde{\rho}(0)\partial_x^2\tilde{\rho}(0):,
\end{align}
and higher order terms associated with $O_{2i\ge6}$ are neglected in the second line of Eq.~\eqref{eq:H+expand}.
\end{subequations}
Here, the expansion coefficients are $c_2=\alpha$ and $c_4=-\alpha^3C(\alpha)$~\cite{note_c4} with 
\begin{align}
C(\alpha)=\frac{\alpha}{24\pi}\left[\frac{\Gamma(\frac{\alpha}{2(1-\alpha)})}{\Gamma(\frac{1}{2(1-\alpha)})}\right]^3\frac{\Gamma(\frac{3}{2(1-\alpha)})}{\Gamma(\frac{3\alpha}{2(1-\alpha)})}.
\end{align}
Specific values of this factor are $C(\alpha\ll1)=1/(4\pi^2\alpha)$, $C(\alpha=1/2)=1/12$, and $C(\alpha=1)=\sqrt{3}/(8\pi)$.

For the heat current between the two heat baths, we move back to the left/right-bath frame from the even/odd-bath frame.
The effective Hamiltonian around the IR fixed point is given by $H_{\rm eff}=H_L^0+H_R^0+V_2+V_4$.
Here, $H_{r}^0$ ($r=L,R$) are the free bosonic Hamiltonians for $\tilde{\phi}_r(x)$ and $\tilde{\rho}_r(x)$ at the IR fixed point and $V_2=c_2 T_K(v/T_K)^2O_2$ and $V_4=c_4T_K(v/T_K)^4O_4$ are the corrections arising in the expansion around the IR fixed point at order $\mathcal{O}(v/T_K)^{n}$.
Note that the lowest anharmonic term is $V_4$ in the absence of the detuning term in the two-level system, whereas $V_3\sim\mathcal{O}(v/T_K)^3$ becomes allowed when the detuning term is present~\cite{Goldstein2013}.
For the later discussion, we decompose the correction as $V_4=V_4^{(a)}+V_4^{(b)}$, where $V_4^{(a)}$ corresponds to the first term of the irrelevant operator $O_4$ in Eq.~\eqref{eq:O4}, and $V_4^{(b)}$ to the second term.

Using the perturbation theory in the corrections $V_2$ and $V_4$ and neglecting contributions of order $\mathcal{O}(T/T_K)^6$ and higher, the steady-state heat current reads $J\approx J_2+J_4^{(a)}+J_4^{(b)}$,
\begin{subequations}
\label{eq:Jdef}
\begin{align}
J_{2}
&=\frac{1}{2}\int_{-\infty}^\infty dt~\braket{[[H_L^0,V_2(0)],V_2(t)]}_0, \\
J_{4}^{(a)}
&=\int_{-\infty}^\infty dt~\braket{[[H_L^0,V_2(0)],V_4^{(a)}(t)]}_0, \\
J_{4}^{(b)}
&=\int_{-\infty}^\infty dt~\braket{[[H_L^0,V_2(0)],V_4^{(b)}(t)]}_0,
\end{align}
\end{subequations}
where $\braket{\cdot}_0$ denotes the quantum mechanical average with respect to the free bosonic Hamiltonian, $H_L^0+H_R^0$.
Each contribution is calculated as (see Appendix~\ref{app:J})
\begin{subequations}
\label{eq:JIR}
\begin{align}
\label{eq:J2}
J_2
&=\frac{2}{15}\pi^3T_K^2\alpha_L\alpha_R(\tilde{T}_L^4-\tilde{T}_R^4), \\
\label{eq:J4}
J_4^{(a)}
&=-\frac{16}{15}\pi^5T_K^2\alpha_L\alpha_R\alpha C(\alpha)(\alpha_L\tilde{T}_L^2+\alpha_R\tilde{T}_R^2)(\tilde{T}_L^4-\tilde{T}_R^4), \\
\label{eq:J3}
J_4^{(b)}
&=\frac{32}{63}\pi^5T_K^2\alpha_L\alpha_R\alpha^2C(\alpha)(\tilde{T}_L^6-\tilde{T}_R^6),
\end{align}
\end{subequations}
where $\tilde{T}_r=T_r/T_K$ is the dimensionless temperature normalized by the Kondo temperature.

The leading contribution $J_2$ reproduces the steady-state heat current~\eqref{eq:Jcotunn} in the weak-coupling regime since $T_K\approx\Delta/\pi$ as $\alpha\to0$.
Then, the low-temperature asymptote of the linear thermal conductance is obtained as
\begin{align}
G_{\rm th}
=\frac{8}{15}\pi^3\alpha_L\alpha_RT_K\left(\frac{T}{T_K}\right)^3.
\end{align}
This low-temperature asymptote exhibits the cubic temperature dependence, which is consistent with thermal conductance in the weak-coupling regime where $T_K(\alpha\ll1)=\Delta/\pi$ (see Eq.~\eqref{eq:Gwc}).

While the leading contribution $J_2$ does not induce heat rectification, the next-leading contribution $J_4^{(a)}$ can give rise to rectification, since it originates from the anharmonic term $\propto\tilde{\rho}^4$ in Eq.~\eqref{eq:O4}.
By contrast, the contribution $J_4^{(b)}$ does not induce rectification either, because it comes from a harmonic term in the field $\propto\tilde{\rho}\partial_x^2\tilde{\rho}$ in Eq.~\eqref{eq:O4}, and thus only renormalizes the heat conductance.
At low temperatures ($T\ll T_K$), the rectification ratio is obtained as
\begin{align}
\label{eq:RIR}
\mathcal{R}\approx1-\delta\tilde{\alpha}|\delta \tilde{T}|(4\pi\alpha)^2C(\alpha)\left(\frac{T}{T_K}\right)^2.
\end{align}
This expression indicates that the rectification ratio is always less than unity, and that its deviation from unity exhibits a nontrivial $\alpha$ dependence.
This behavior is a signature of quantum many-body effects.
As the temperature is lowered, the deviation is quadratically suppressed, and heat rectification asymptotically vanishes.

\begin{figure}
    \centering
    \includegraphics[width=1\linewidth]{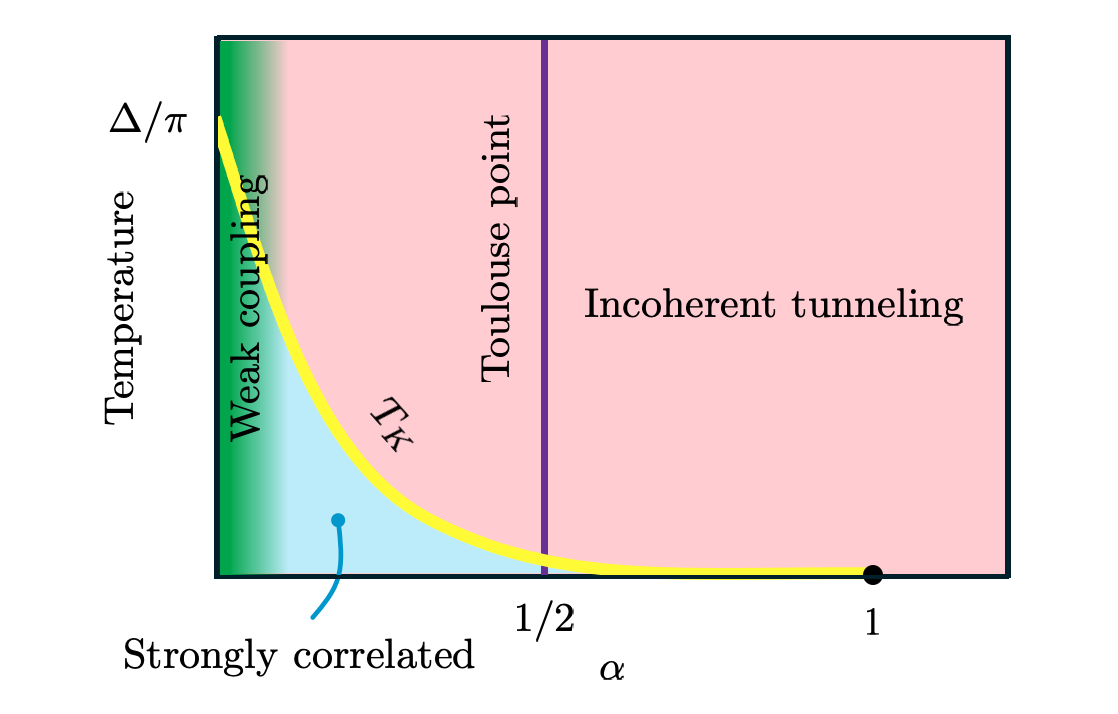}
    \caption{Schematic overview of the regimes considered in this work: the weak-coupling regime, the incoherent-tunneling regime, the Toulouse point, and the strongly correlated regime. For $\alpha<1$, the strongly correlated regime and the incoherent-tunneling regime are separated by the Kondo temperature $T_K$, which decreases from $\Delta/\pi$ at $\alpha=0$ and vanishes at $\alpha=1$. By contrast, for $\alpha>1$, the incoherent-tunneling regime extends over the entire temperature range.}
    \label{fig:TK}
\end{figure}

We summarize the regimes in which the analytical formulas were derived in the parameter space of the coupling strength $\alpha$ and temperature, as shown in Fig.~\ref{fig:TK}.

\section{Numerical results}
\label{sec:numerics}

In this section, we show the numerical results and compare them with the analytical formulas derived in Sec.~\ref{sec:analytics} for the linear thermal conductance and the rectification ratio by calculating the steady-state heat current after briefly explaining the numerical procedure for calculating them.

\subsection{Numerical procedure}

In this work, we calculate the real-time dynamics of the correlation function $\chi(t,t')$ using a tensor network approach, the TEMPO method~\cite{Strathearn2018} as follows.
First, we evolve the reduced density matrix of the two-level system $\rho_{\rm TLS}(t')=U(t',0)\rho_{\rm TLS}(0)U^\dagger(t',0)$ from the initial state up to a sufficiently long time $t'$, by which the system has reached the steady state.
We then apply $\sigma_z$ from the right and evolve $\rho_{\rm TLS}(t')\sigma_z$ further up to time $t$ using the same procedure as before $t'$.
Finally, we obtain the correlation function from $\chi(t-t')=\theta(t-t'){\rm Im}[{\rm tr}[U(t,t')\rho_{\rm TLS}(t')\sigma_zU^\dagger(t,t')]]$ (see Appendix~\ref{app:tempo} for details).
We then evaluate the steady-state heat current from the Meir–Wingreen formula~\eqref{eq:J}.

\subsection{Linear thermal conductance}

\begin{figure*}[t]
  \centering
  \includegraphics[width=1.0\textwidth]{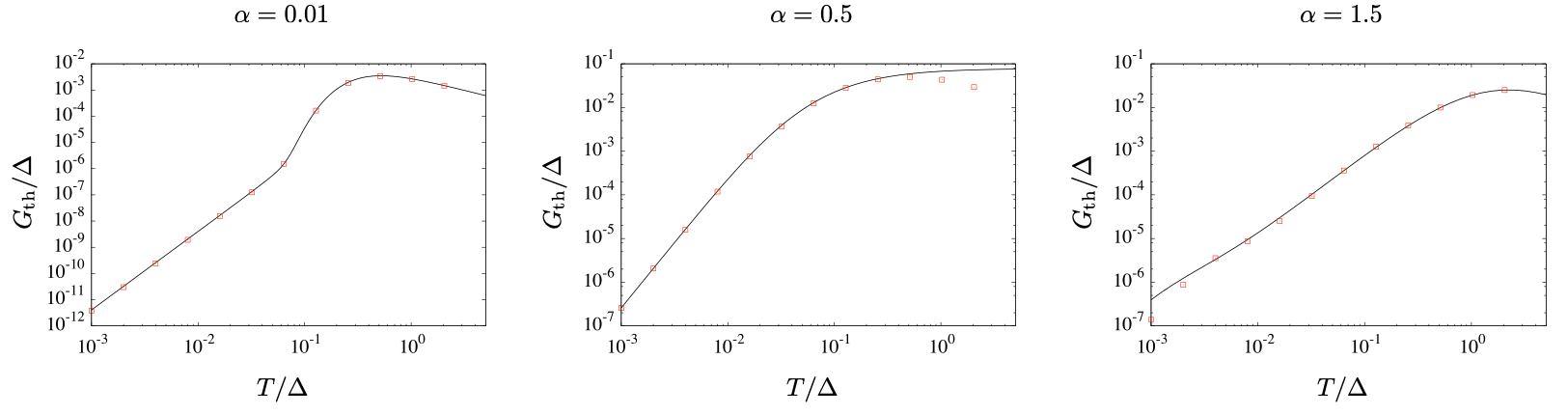}
  \caption{Linear thermal conductance $G_{\rm th}$ as a function of the average temperature $T$ for different symmetric couplings $\alpha_L=\alpha_R=\alpha/2$, with $|\delta T|=0.001T$ and $\omega_c=5\Delta$. The plots represent numerical simulations by the TEMPO algorithm, while the solid lines correspond to analytical formulas in each coupling regime: perturbation theory in the system-bath coupling~\eqref{eq:Jwc}, the exact solution at the Toulouse point~\eqref{eq:chi_toulouse}, and the NIBA~\eqref{eq:Jniba} for $\alpha=0.01$ (left panel), 0.5 (middle panel), and 1.5 (right panel), respectively. We use the cutoff frequency $\Omega_{\rm T}=1.25\omega_c$ for plotting the analytical formula at the Toulouse point.}
  \label{fig:Gth}
\end{figure*}

We first compare the numerical results for the linear thermal conductance calculated by the TEMPO algorithm with the analytical formulas in different system-bath coupling regimes for the symmetric coupling case ($\alpha_L=\alpha_R=\alpha/2$) to confirm the numerical accuracy.
The linear thermal conductance is calculated from the definition, $G_{\rm th}=J/\delta T$, by applying the small temperature difference $|\delta T|=0.001 T$.
Figure~\ref{fig:Gth} shows the temperature dependence of the linear thermal conductance for different coupling strengths, $\alpha=0.01$, 0.5, and 1.5.

For the weak-coupling regime ($\alpha=0.01$), the perturbation theory in the system-bath coupling~\eqref{eq:Jwc} (solid line in the left panel) describes heat transport very well.
The crossover between the sequential tunneling and cotunneling processes is estimated as $T=0.1\Delta$.
At low temperatures ($T\lesssim0.1\Delta$), the linear thermal conductance shows the cubic temperature dependence (see Eq.~\eqref{eq:Gwc}).

At the Toulouse point ($\alpha=1/2$), the numerical results are also in good agreement with the exact solution~\eqref{eq:chi_toulouse} (solid line in the middle panel) except for the high temperature, $T\gtrsim\Delta$.
The deviation at high temperature is attributed to the high-frequency cutoff effects.
We note that the cutoff frequencies $\omega_c$ and $\Omega_{\rm T}$, introduced in the Ohmic spin-boson model and the noninteracting resonant-level model, respectively, are on the same scale, but are not identical.
By comparing the numerical results based on the Ohmic spin-boson model with the exact low-temperature solution obtained from the noninteracting resonant-level model, we find $\Omega_{\rm T}\simeq 1.25\,\omega_c$.

For the strong coupling ($\alpha=1.5$), the NIBA results~\eqref{eq:Jniba} as a solid line in the right panel match the numerical results over the entire temperature range.
At low temperatures ($T\lesssim\omega_c$), the linear thermal conductance exhibits a power-law dependence, $G_{\rm th}\propto T^{2\alpha-1}$ as predicted in Sec.~\ref{sec:niba}.
The small fluctuations in the low-temperature numerical data can be attributed to the long relaxation time, $\tau_{\rm rel}\sim(2\pi\alpha T)^{-1}$, which makes it difficult to reach the steady state for the low temperature.

\subsection{Rectification ratio}

\begin{figure}[b]
    \centering
    \includegraphics[width=1.0\linewidth]{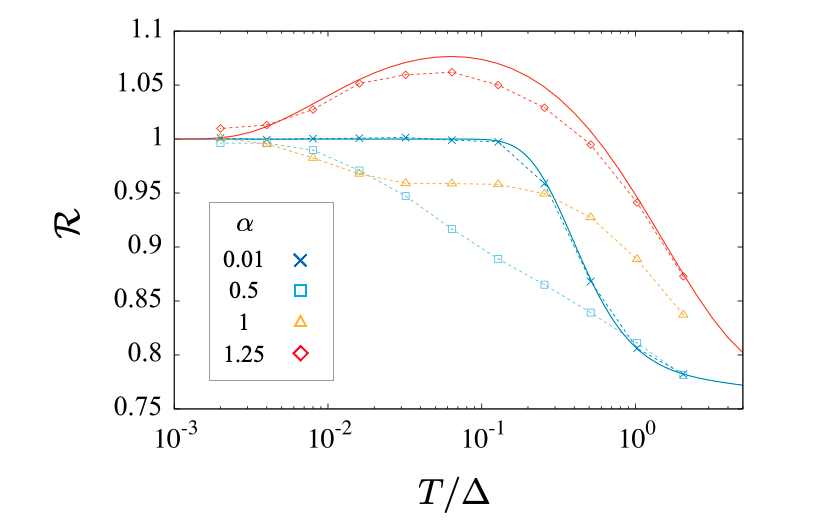}
    \caption{Heat rectification ratio $\mathcal{R}$ as a function of the average temperature $T$ for different coupling strengths $\alpha$ and $\omega_c=5\Delta$. The plots denote the numerical results obtained by the TEMPO algorithm, and the blue and red solid lines represent the analytical formulas for the weak-coupling and the incoherent-tunneling regimes, respectively. The asymmetric parameters are set to $\delta\alpha=\alpha_L-\alpha_R=0.5\alpha$ and $|\delta T|=|T_L-T_R|=0.5T$.}
    \label{fig:R_vs_T}
\end{figure}

Next, we present the numerical results for the rectification ratio, defined as~\eqref{eq:R_def}.
We first show the temperature dependence of the rectification ratio for different coupling strengths $\alpha$ in Fig.~\ref{fig:R_vs_T}.
For $\alpha\lesssim1$, the rectification ratio decreases monotonically from $\mathcal{R}=1$ as the temperature increases, thus remains smaller than unity over the entire temperature range.
For the weak coupling ($\alpha=0.01$), the numerical results are in good agreement with the analytical result weak-coupling regime~\eqref{eq:Jwc}, shown as the blue solid line in Fig.~\ref{fig:R_vs_T}.
As the average temperature $T=(T_L+T_R)/2$ is increased while keeping the temperature ratio $T_L/T_R$ fixed, the rectification ratio starts to decrease from $\mathcal{R}=1$ around $T\sim0.1\Delta$, and then approaches the high-temperature limiting value~\eqref{eq:Rwc}, $\mathcal{R}\approx 0.778$ for $\delta\alpha=0.5\alpha$ and $\delta T=0.5T$.

As the system-bath coupling increases from the weak-coupling regime, the rectification ratio first decreases.
However, for the stronger coupling, it turns out to increase.
Eventually, for the strong system-bath coupling ($\alpha=1.25$), the rectification ratio exceeds unity at low temperatures ($T\lesssim0.1\omega_c$), but rectification vanishes ($\mathcal{R}\to1$) as $T\to0$, as expected in Sec.~\ref{sec:niba}.
In this region, the NIBA is valid over the entire temperature range and qualitatively agrees with the numerical results, as shown by the red line in Fig~\ref{fig:R_vs_T}.

\begin{figure}
    \centering
    \includegraphics[width=1.0\linewidth]{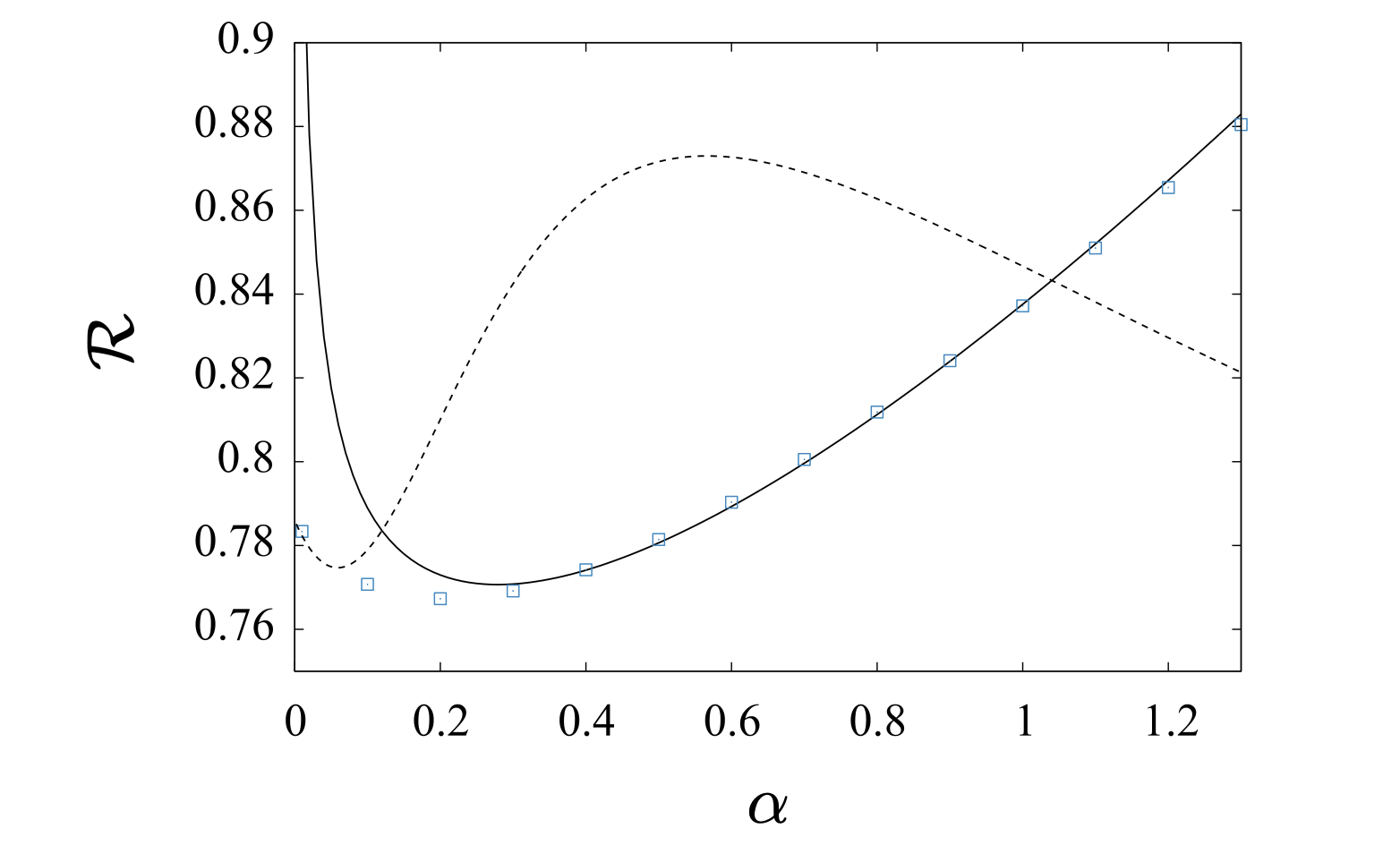}
    \caption{Heat rectification ratio $\mathcal{R}$ as a function of the total coupling strength $\alpha$ for the high temperature regime, $T=2.048\Delta$. Plots represent the numerical results obtained using the TEMPO algorithm, and the solid and dotted lines denote the NIBA and the weak-coupling, respectively. The other parameters are the same as those in Fig~\ref{fig:R_vs_T}.}
    \label{fig:R_vs_alpha}
\end{figure}

To illustrate the crossover from the weak-coupling to the strong-coupling regimes, we show in Fig.~\ref{fig:R_vs_alpha} the dependence of the rectification ratio on the total coupling strength $\alpha$ in the high-temperature regime ($T=2.048\Delta$).
We note that the rectification ratio remains nearly unity at low temperatures for arbitrary coupling strengths.
The rectification ratio exhibits nonmonotonic behavior. The NIBA results agree well with the numerical results over a wide range of coupling strengths $\alpha$.
However, as the coupling strength is reduced to $\alpha=0.2$, the numerical results begin to deviate from the NIBA results and enter the weak-coupling regime.

Finally, we turn to the strongly correlated regime.
Since the rectification ratio is close to unity in this regime, we plot its negative deviation from unity, $1-\mathcal{R}$.
Figure~\ref{fig:R_IR} shows the temperature dependence of this deviation for several values of the coupling strength.
At low temperatures ($T\ll T_K$), the numerical results for $1-\mathcal{R}$ show a quadratic suppression, in agreement with the prediction of the effective model near the IR fixed point~\eqref{eq:RIR}.
With increasing temperature, the numerical data deviates from the quadratic temperature dependence.
The onset of this deviation shifts to lower temperatures as the coupling strength increases, reflecting the $\alpha$ dependence of the Kondo temperature (see Fig.~\ref{fig:TK}).
We note that the cutoff frequency entering the Kondo temperature~\eqref{eq:TK} may depend on details of the underlying model.
To obtain consistency with the numerical results, we use $1.75\,\omega_c$ as a cutoff frequency entering the Kondo temperature in Eq.~\eqref{eq:RIR}, shown as solid lines in Fig.~\ref{fig:R_IR}. 

\begin{figure}[t]
    \centering
    \includegraphics[width=1.0\linewidth]{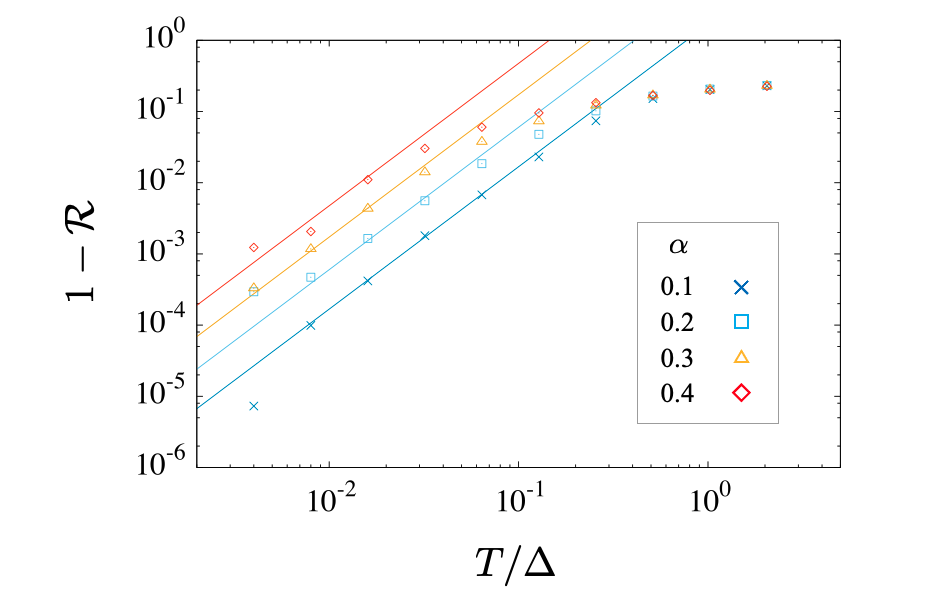}
    \caption{Temperature dependence of $1-\mathcal{R}$, the deviation of the rectification ratio from unity for different coupling strengths. The solid lines represent the analytical formulas~\eqref{eq:RIR} from the effective model near the IR fixed point. The other parameters are the same as those in Fig.~\ref{fig:R_vs_T}.}
    \label{fig:R_IR}
\end{figure}

\section{Summary}
\label{sec:summary}

We have studied quantum heat transport through a two-level system asymmetrically coupled to two thermal baths.
This system is described by the Ohmic spin-boson model and can be equivalently mapped onto the anisotropic Kondo model.
We evaluated the steady-state heat current using the TEMPO algorithm by computing the correlation function entering the exact heat-current formula, and compared the numerical results with analytical formulas in several limiting regimes.

First, we confirmed that the numerical results for the linear thermal conductance agree with the analytical formulas in the weak-coupling regime ($\alpha\ll1$), the incoherent-tunneling regime ($T>T_K$), and the Toulouse point ($\alpha=1/2$).
These comparisons validate the tensor-network approach over a broad range of coupling strengths and temperatures.

We then investigated the rectification ratio of the steady-state heat currents obtained by exchanging the temperatures of the two thermal baths.
By comparing the numerical results with analytical formulas, we found a non-unity rectification ratio in several regimes: $\mathcal{R}<1$ at high temperatures in the weak-coupling and the incoherent-tunneling regimes, and $\mathcal{R}>1$ at lower temperatures in the incoherent-tunneling regime.
We also numerically observed the crossover from the weak-coupling regime to the incoherent-tunneling regime at high temperatures.

Finally, we focused on the strongly correlated regime ($T<T_K$).
In this regime, the rectification ratio remains close to unity, but its deviation from unity increases quadratically with temperature, as predicted by perturbation theory near the infrared fixed point.
Our numerical results agree well with this analytical prediction.
As the temperature increases, the numerical data start to deviate from this quadratic behavior, signaling the crossover out of the perturbative regime near the infrared fixed point.
These results show that heat-rectification effects, even in such a minimal nonlinear system, encode signatures of dissipation-induced many-body physics and provide a basis for understanding nonlinear quantum heat transport in strongly dissipative quantum thermal devices.

\section*{Acknowledgments}

We acknowledge funding through the project T-KONDO ANR-24-CE30-0262, and TY was supported by JSPS Overseas Research Fellowships.

\appendix

\section{Numerical method}
\label{app:tempo}

In this Appendix, we describe the details for calculating the heat current using the tensor network simulation following Ref.~\cite{Strathearn2018}.

\subsection{Feynman-Vernon path integral}

First, we show the real-time dynamics of the two-level system using the Feynman-Vernon path-integral formalism~\cite{Feynman1963}.
The time-evolution of the reduced density matrix of the two-level system, $\rho(t)={\rm tr}_{\rm B}[\varrho(t)]$, where $\varrho(t)$ is the density matrix of the global system and ${\rm tr}_{\rm B}[O]$ denotes tracing over with respect to the degree of freedom of the baths, is given by~\cite{Leggett1987, Weiss_text}
\begin{align}
[\rho]^{y_M}=\sum_{\sigma_0,\dots,\sigma_M}\prod_{n=0}^{M-1}w(y_{n+1},y_n)\prod_{m\le n=0}^Me^{-\phi(y_n,y_m)}[\rho]^{y_0},
\end{align}
where we introduce the vectorized index combined two eigenvalues of $\sigma_z$, $y=(\sigma,\sigma')$, and thus the index takes four values, $y=0$ for ($\sigma=+1$, $\sigma'=+1$), $y=1$ for ($\sigma=+1$, $\sigma'=-1$), $y=2$ for ($\sigma=-1$, $\sigma'=+1$), and $y=3$ for ($\sigma=-1$, $\sigma'=-1$).
Using the vectorized index, we can express the matrix element of the reduced density matrix as a tensor form, $[\rho]^{y_n}=[\rho]^{(\sigma_n,\sigma_n')}=\bra{\sigma_n}\rho(t_n)\ket{\sigma_n'}$ at the discretized time $t_n=n\delta t_n$ ($n=0,\dots,M$) with $\delta t_{0<n<M}=\delta t$ and $\delta t_0=\delta t_M=\delta t/2$.
In this work, we prepare the excited state as an initial state at $t_0$, i.e., $[\rho]^{y_0}=\braket{\sigma_0|+1}\braket{+1|\sigma_0'}=\delta_{y_0,0}$.

The time evolution of the reduced density matrix is influenced by the two-level system itself and the coupling to the baths.
The time evolution by the Hamiltonian of the two-level system is described by
\begin{align}
w(y_{n+1},y_n)
&=\bra{\sigma_{n+1}}U(t_{n+1};t_n)\ket{\sigma_n} \nonumber\\
&\quad~\times\bra{\sigma'_n}U(t_{n};t_{n+1})\ket{\sigma_{n+1}'},
\end{align}
where $U(t_{n+1},t_n)$ is the time-evolution operator of the two-level system by the small time step $t_{n+1}-t_n=\delta t$,
\begin{align}
U(t_{n+1};t_n)=e^{-i\delta t\Delta\sigma_x/2}+\mathcal{O}(\delta t)^3.
\end{align}
The influence for the time evolution by the coupling to the baths is incorporated in the influence phase,
\begin{align}
\phi(y_n,y_m)=(\sigma_n-\sigma_n')(\eta_{n,m}\sigma_m-\eta_{n,m}^*\sigma_m'),
\end{align}
where $\eta_{n,m}$ describes the memory effect in non-Markovian dynamics~\cite{Makri1995I, Makri1995II},
\begin{subequations}
\begin{align}
\eta_{n>m}&=\delta t_n\delta t_m\sum_{r=L,R}K_r(t_n-t_m), \\
\eta_{n=m}&=(\delta t_n)^2\sum_{r=L,R}\frac{K_r(0)}{2}.
\end{align}    
\end{subequations}
Here, $K_r(t)=K_r'(t)+iK''_r(t)$ is the autocorrelation function of the bath $r$,
\begin{subequations}
\begin{align}
K'_r(t)
&=\frac{\alpha_r}{2}\int_0^\infty d\omega~\omega e^{-\omega/\omega_c}\coth\left(\frac{\omega}{2T_r}\right)\cos(\omega t), \\
K''_r(t)
&=-\frac{\alpha_r}{2}\int_0^\infty d\omega~\omega e^{-\omega/\omega_c}\sin(\omega t).
\end{align}
\end{subequations}
The influence phase $\phi(y_n,y_m)$ depends on the past history, while $w(y_{n+1},y_n)$ depends only on the immediate previous state.

\subsection{TEMPO}

The memory effects caused by the coupling to the bath prevent us from computing the long-time dynamics.
To overcome this, we adopt the tensor network algorithm, the time-evolving matrix product operators (TEMPO)~\cite{Strathearn2018}. 

Introducing the augmented density tensor (ADT)
\begin{align}
A^{y_0,\dots,y_M}=\prod_{m\le n=0}^M \mathcal{F}^{y_n,y_m} [\rho]^{y_0},
\end{align}
where $\mathcal{F}^{y_n,y_m}$ is the renormalized influence functional,
\begin{align}
\mathcal{F}^{y_n,y_m}=
\begin{dcases}
e^{-\phi(y_{n},y_{n-1})}w(y_{n},y_{n-1}), & (n=m+1>0), \\
e^{-\phi(y_n,y_m)}, & ({\rm otherwise}),
\end{dcases}
\end{align}
the reduced density matrix is expressed by summing over all indices except for the latest time step,
\begin{align}
[\rho]^{y_M}=\sum_{y_0,\dots,y_{M-1}}A^{y_0,\dots,y_{M-1},y_M}.
\end{align}
Thus, from this relation, we can compute the reduced density matrix at $t_M$ from the $(M+1)$-leg ADT $A^{y_0,\dots,y_{M}}$. 
The ADT evolves step by step through the transfer tensor $B$ as
\begin{align}
A^{y_0,\dots,y_{n},y_{n+1}}
&=B^{y_0,\dots,y_{n},y_{n+1}}_{y'_0,\dots,y_n'}A^{y'_0,\dots,y'_{n}}, \\
B^{y_0,\dots,y_{n},y_{n+1}}_{y'_0,\dots,y_n'}
&=\prod_{m=0}^n\delta_{y_m'}^{y_m}\times\prod_{m=0}^{n+1}\mathcal{F}^{y_n,y_m}.
\end{align}

\begin{figure}
    \centering
    \includegraphics[width=1.0\linewidth]{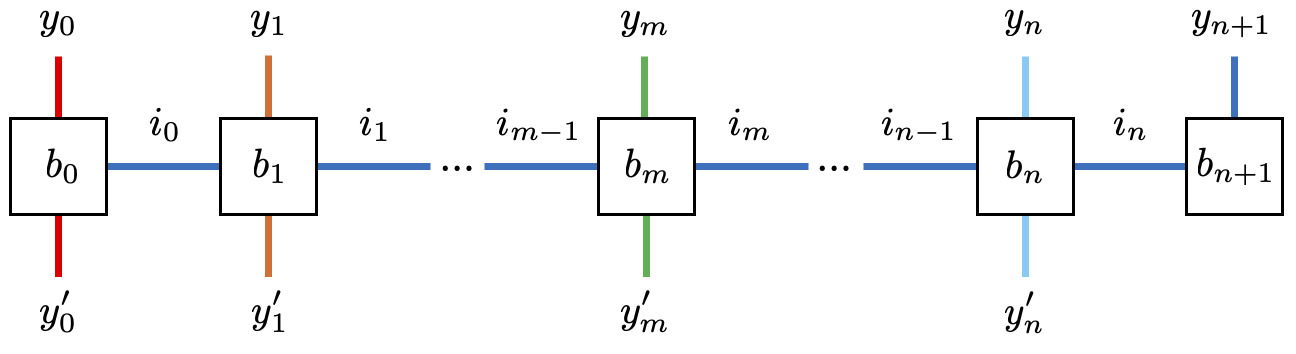}
    \caption{MPO representation of the transfer matrix $B$ with $(2n+1)$ legs. Legs of the same color carry the same information via the Kronecker delta function, e.g., $\delta_{y_0'}^{y_0}$.}
    \label{fig:mpo}
\end{figure}

For the efficient calculation of the tensor multiplication using the MPS compression algorithm, we represent the ADT and the transfer tensor as a matrix product state (MPS) and the matrix product operator (MPO), respectively.
The MPO representation of the transfer matrix is (see the tensor network diagram in Fig.~\ref{eq:J}) 
\begin{align}
B^{y_0,\dots,y_{n},y_{n+1}}_{y'_0,\dots,y_n'}
=[b_0]_{y'_0}^{i_0,y_0}\left(\prod_{m=1}^n[b_m]_{i_{m-1},y'_m}^{i_m,y_m}\right)[b_{n+1}]_{i_n}^{y_{n+1}},
\end{align}
where the 4-leg tensor in the middle of the MPO is
\begin{align}
[b_m]_{i_{m-1},y'_m}^{i_m,y_m}=\delta_{i_{m-1}}^{i_m}\delta_{y_m'}^{y_m}\mathcal{F}^{i_m,y_m},
\end{align}
the 3-leg tensor at the left edge of the MPO is
\begin{align}
[b_0]_{y_0'}^{i_0,y_0}=\delta_{y_0}^{y_0'}\mathcal{F}^{i_0,y_0},
\end{align}
and the 2-leg tensor at the right edge of the MPO is
\begin{align}
[b_{n+1}]_{i_n}^{y_{n+1}}=\delta_{i_n}^{y_{n+1}}\mathcal{F}^{y_{n+1},y_{n+1}}.
\end{align}
By applying this MPO $B$ with $(2n+1)$ legs to the MPS $A$ with $n$ legs, we obtain the ADT with $(n+1)$ legs.
The resulting ADT is then represented as an MPS by performing a singular value decomposition.
In this work, we perform the above procedure using \texttt{applyMPO(MPO A, MPS x)} based on the density matrix MPS compression algorithm implemented in the \texttt{ITensor} library~\cite{Fishman2022}.
The truncation error cutoff is set to $10^{-16}$, and the maximum bond dimension is $100$ for the figures in the main text.

Moreover, we truncate memory contributions beyond $\tau_c = n_c \delta t$ so that the number of tensors in the MPS of the ADT remains $n_c+1$.
To this end, for $n>n_c$, we construct a $2(n_c+1)$-leg transfer tensor by summing over the earliest time index $y_{n-n_c}$,
\begin{align}
\tilde{B}^{y_{n-n_c+1},\dots,y_{n},y_{n+1}}_{y'_{n-n_c},\dots,y_n'}
&=\sum_{y_{n-n_{\rm c}}}[b_{n-n_c}]_{y'_{n-n_c}}^{i_{n-n_c},y_{n-n_c}} \nonumber\\
&\times\left(\prod_{m=n-n_c+1}^n[b_m]_{i_{m-1},y'_m}^{i_m,y_m}\right)[b_{n+1}]_{i_n}^{y_{n+1}}.
\end{align}
Applying this MPO to the ADT yields a $(n_c+1)$-leg MPS representing the ADT at $t=t_{n+1}$.

\subsection{Steady-state heat current}

The steady-state heat current~\eqref{eq:J} is rewritten as
\begin{align}
J=\frac{\alpha_L\alpha_R}{\alpha}\int_0^\infty dt~\chi(t)[F(t,T_L)-F(t,T_R)],
\end{align}
where $F(t,T_r)$ is the window function 
\begin{align}
F(t,T_r)
&=\int_0^\infty d\omega~\omega^2e^{-\omega/\omega_c}\sin(\omega t)n_r(\omega) \nonumber\\
&=T_r^3{\rm Im}\left[\psi^{(2)}(1+(it+\omega_c^{-1})T_r)\right],
\end{align}
where $\psi^{(2)}(z)=d^2\psi(z)/dz^2$ is the polygamma function of order 2, and it decays as $F(t,T_L)-F(t,T_R)\sim2\omega_c^2(T_L-T_R)/(\omega_ct)^{3}$ for large $t$.
Therefore, we can evaluate the steady-state heat current by tracing the real-time dynamics of the correlation function,
\begin{align}
\chi(t,t_0)
&=i\theta(t-t_0)[\braket{\mathcal{U}^\dagger(t)\sigma_z\mathcal{U}(t,t_0)\sigma_z\mathcal{U}(t_0)} \nonumber\\
&\qquad\qquad\quad-\braket{\mathcal{U}^\dagger(t_0)\sigma_z\mathcal{U}^\dagger(t,t_0)\sigma_z\mathcal{U}(t)}] \nonumber\\
&=\theta(t-t_0){\rm Im}\left[{\rm tr}[\mathcal{U}(t,t_0)\varrho(t_0)\sigma_z\mathcal{U}^\dagger(t,t_0)\sigma_z]\right],
\end{align}
where $\mathcal{U}(t,t_0)$ is the time-evolution operator of the total system.
We show the real-time dynamics of the correlation function for symmetric coupling ($\alpha_L=\alpha_R=\alpha/2$) in Fig.~\ref{fig:chi}, using $\delta t=0.1\Delta^{-1}$, $t_0=200\Delta^{-1}$, and $\tau_c=100\Delta^{-1}$, which are the numerical parameters used for the figures in the main text.
This expression indicates that the correlation function can be computed as follows: first, we evolve the reduced density matrix up to time $t_0$ using the TEMPO algorithm, next, we apply $\sigma_z$ from the right, then we further evolve $\rho(t_0)\sigma_z$ again in the same way as up to $t_0$, and finally, we take the quantum mechanical expectation value of $\sigma_z$ with respect to $U(t,t_0)\rho(t_0)\sigma_z U^\dagger(t,t_0)$.

\begin{figure}
    \centering
    \includegraphics[width=1.0\linewidth]{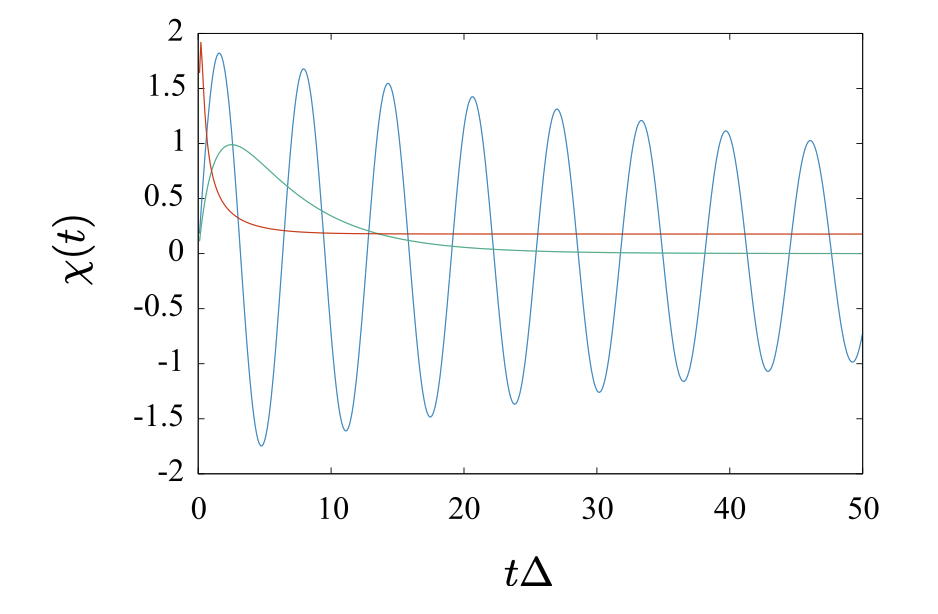}
    \caption{Time evolution of the correlation function $\chi(t)$ for $\alpha=0.01$ (blue), 0.5 (green), 1.5 (red), with $T=0.032\Delta$, $|\delta T|=0.001T$, and $\omega_c=5\Delta$. The curve for $\alpha=1.5$ is scaled by a factor of 100.}
    \label{fig:chi}
\end{figure}

\section{Heat current around IR fixed point}
\label{app:J}

We derive the steady-state heat current using the effective Hamiltonian around the IR fixed point.

The heat current operator flowing out of the left bath is defined as
\begin{align}
\hat{J}\equiv-\frac{dH_L^0}{dt}=\sum_{n=2,4}i[H_L^0,V_n],
\end{align}
where $V_4=V_4^{(a)}+V_4^{(b)}$.
The perturbation theory in the correction provides the heat current,
\begin{align}
J
&=\braket{\hat{J}}_0-i\sum_{n}\int_{-\infty}^0 dt~\braket{[\hat{J}(0),V_n(t)]}_0 \nonumber\\
&=\braket{\hat{J}}_0+\sum_{n,n'}J_{n,n'},
\end{align}
where
\begin{align}
J_{n,n'}=\int_{-\infty}^0 dt~\braket{[[H_L^0,V_n(0)],V_{n'}(t)]}_0.
\end{align}
Here, $\braket{\cdot}_0$ is the quantum mechanical average with respect to the free bosonic Hamiltonian $H_L^0+H_R^0$. 
Now, since $V_n$ is of order $\mathcal{O}(1/T_K)^{n-1}$, the steady-state heat current can be expanded as a series in powers of $\mathcal{O}(1/T_K)^{n+n'-2}$, such as $\mathcal{O}(1/T_K)^2$, $\mathcal{O}(1/T_K)^4$, and $\mathcal{O}(1/T_K)^6$.
By neglecting terms of order $\mathcal{O}(1/T_K)^{6}$ and higher, we are only left with $J_{2,2}\sim\mathcal{O}(1/T_K)^2$ and $J_{2,4}\sim J_{4,2}\sim\mathcal{O}(1/T_K)^4$.

For the steady-state, the zeroth-order contribution vanishes, $\braket{\hat{J}}_0=0$.
Using $\braket{[[H_L^0,V_n(0)],V_{n'}(t)]}_0=\braket{[[H_L^0,V_{n'}(0)],V_{n}(-t)]}_0$, we obtain
\begin{align}
J_{n,n'}+J_{n',n}=\int_{-\infty}^{\infty} dt~\braket{[[H_0^L,V_n(0)],V_{n'}(t)]}_0,
\end{align}
and thus the steady-state heat current is written as
\begin{align}
J\approx J_2+J_4^{(a)}+J_4^{(b)},
\end{align}
where $J_2$, $J_4^{(a)}$ and $J_4^{(b)}$ are defined in Eq.~\eqref{eq:Jdef}.

\subsection{Contribution $J_2$}

The commutation $[H_0^L,V_2]$ is
\begin{align}
[H_L^0,V_2]
&=-ig_2[\alpha_1(:\rho_L\dot{\rho}_L:+:\dot{\rho}_L\rho_L:) \nonumber\\
&\qquad\qquad+2\sqrt{\alpha_L\alpha_R}:\dot{\rho}_L\rho_R:],
\end{align}
where we write $\rho_r=\tilde{\rho}_r(x=0)$ for simplicity and $g_2=\pi c_2T_K/\alpha(v/T_K)^2$.
Noting that, for the free bosonic Hamiltonian, the different modes are separable, $\braket{O_LO_R}_0=\braket{O_L}_L\braket{O_R}_R$ and the odd-order moments are zero, $\braket{\rho_r^{2n+1}}_r=0$, the steady-state heat current reads
\begin{align}
J_2
&=-4i\alpha_L\alpha_Rg_2^2 \nonumber\\
&\quad~\times\int_{-\infty}^\infty dt~\braket{[:\dot{\rho}_L(0)\rho_R(0):,:\rho_L(t)\rho_R(t):]}_0.
\end{align}
We note that $\int dt~\braket{[\rho_L(t)\dot{\rho}_L(0)+\dot{\rho}_L(t)\rho_L(0),\rho_r(t)]}_0=0$
Using the Wick's theorem, the heat current is expressed as
\begin{align}
J_2
=4ig_2^2\alpha_L\alpha_R\int_{-\infty}^\infty dt~\left[\dot{\bar{S}}_L(t)\bar{S}_R(t)-\dot{S}_L(t)S_R(t)\right],
\end{align}
where $S_r(t)=\braket{\rho_r(t)\rho_r(0)}_r$ and $\bar{S}_r(t)=\braket{\rho_r(0)\rho_r(t)}_r$ are auto-correlation functions, and then, in the Fourier space, it is
\begin{align}
J_2&=-4g_2^2\alpha_L\alpha_R \nonumber\\
&\quad~\times\int_{-\infty}^\infty \frac{d\omega}{2\pi}~\omega\left[S_L(\omega)S_R(-\omega)-S_L(-\omega)S_R(\omega)\right],
\end{align}
where $\bar{S}_r(\omega)=S_r(-\omega)$.
Assuming that the baths are in thermal equilibrium, the auto-correlation function holds $S_r(-\omega)=e^{-\omega/T_r}S_r(\omega)$.
Therefore, introducing the spectral function $A(t)=\braket{[\rho_r(t),\rho_r(0)]}_r=S_r(t)-\bar{S}_r(t)$, the steady-state heat current is obtained as
\begin{align}
\label{eq:J2app}
J_2
=8g_2^2\alpha_L\alpha_R\int_0^\infty \frac{d\omega}{2\pi}~\omega A_L(\omega)A_R(\omega)[n_L(\omega)-n_R(\omega)].
\end{align}

\subsection{Contribution $J_4^{(a)}$}

The steady-state heat current $J_4^{(a)}$ is obtained, in the same manner as $J_2$, as
\begin{align}
\label{eq:J4app}
J_4^{(a)}
&=-8ig_2g_4\alpha_L\alpha_R \nonumber\\
&\quad~\times\int_{-\infty}^\infty dt~\Big\{\alpha_L\braket{[:\dot{\rho}_L(0)\rho_R(0):,:\rho_L^3(t)\rho_R(t):]}_0 \nonumber\\
&\qquad\qquad\quad~~+\alpha_R\braket{[:\dot{\rho}_L(0)\rho_R(0):,:\rho_L(t)\rho_R^3(t):]}_0 \nonumber\\
&=24ig_2g_4\alpha_L\alpha_R(\alpha_L\braket{:\rho_L^2:}_L+\alpha_R\braket{:\rho_R^2:}_R) \nonumber\\
&\quad~\times\int_{-\infty}^\infty dt~\left[\dot{\bar{S}}_L(t)\bar{S}_R(t)-\dot{S}_L(t)S_R(t)\right] \nonumber\\
&=48g_2g_4\alpha_L\alpha_R(\alpha_L\braket{:\rho_L^2:}_L+\alpha_R\braket{:\rho_R^2:}_R) \nonumber\\
&\quad~\times\int_0^\infty \frac{d\omega}{2\pi}~\omega A_L(\omega)A_R(\omega)[n_L(\omega)-n_R(\omega)],
\end{align}
where $g_4=\pi^3c_4T_K/\alpha^2(v/T_K)^4$.

\subsection{Contribution $J_4^{(b)}$}

The contribution from $V_4^{(b)}$ is calculated as 
\begin{align}
J_4^{(b)}
&=-2ig_2g_3\alpha_L\alpha_R \nonumber\\
&\quad~\times\int_{-\infty}^\infty dt~\Big\{\braket{
[:\dot{\rho_L}(0)\rho_R(0):,;\rho_L(t)\partial_x^2\rho_R(t):]}_0 \nonumber\\
&\qquad\qquad\quad~+\braket{[:\dot{\rho_L}(0)\rho_R(0):,:\partial_x^2\rho_L(t)\rho_R(t):]}_0 \nonumber\\
&=-2ig_2g_3\alpha_L\sum_{r}\alpha_r\int_{-\infty}^\infty dt~[\dot{S}_L(t)D_r(t)+\dot{D}_L(t)S_r(t) \nonumber\\
&\qquad\qquad\qquad\qquad\qquad\quad~
-\dot{\bar{S}}_L(t)\bar{D}_r(t)-\dot{\bar{D}}_L(t)\bar{S}_r(t)],
\end{align}
with $g_3=-\pi c_4T_K/(2\alpha)(v/T_K)^4$ and the correlation functions, $D_r(t)=\braket{\partial_x^2\rho_r(t)\rho_r(0)}_r$ and $\bar{D}_r(t)=\braket{\rho_r(0)\partial_x^2\rho_r(t)}$.
In the Fourier space, it reads
\begin{align}
J_4^{(b)}
&=-4g_2g_3\alpha_L\alpha_R \nonumber\\
&\quad~\times\int_{-\infty}^\infty \frac{d\omega}{2\pi}~\omega\Big\{S_L(\omega)D_R(-\omega)-D_L(-\omega)S_R(\omega)\Big\},
\end{align}
where we used $\bar{D}_r(\omega)=D_r(-\omega)$.
Since $\partial_x^2\rho=v^{-2}\partial_t^2\rho$, we obtain $D_r(\omega)=-(\omega/v)^2S_r(\omega)$, and thus the heat current is
\begin{align}
\label{eq:J3app}
J_4^{(b)}
&=4g_2g_3\alpha_L\alpha_R\frac{1}{v^2} \nonumber\\
&\quad~\times\int_{-\infty}^\infty \frac{d\omega}{2\pi}~\omega^3[S_L(\omega)S_2(-\omega)-S_L(-\omega)S_R(\omega)] \nonumber\\
&=-8g_2g_3\alpha_L\alpha_R\frac{1}{v^3} \nonumber\\
&\quad~\times\int_0^\infty \frac{d\omega}{2\pi}~\omega^3A_L(\omega)A_R(\omega)[n_L(\omega)-n_R(\omega)].
\end{align}

\subsection{Spectral function}

Now, we calculate the spectral function $A_r(\omega)$.
The bosonic fields in the free bosonic Hamiltonian at the IR fixed point with the length $L$ are expressed as
\begin{subequations}
\begin{align}
\tilde{\phi}_r(x)
&=\sum_{n>0}\sqrt{\frac{\pi}{k_nL}}\cos(k_nx)(a_{k_n}+a_{k_n}^\dagger), \\
\tilde{\rho}_r(x)
&=\sum_{n>0}\frac{1}{i}\sqrt{\frac{k_n}{\pi L}}\cos(k_nx)(a_{k_n}-a_{k_n}^\dagger),
\end{align}
\end{subequations}
where $k_n=n\pi/L$.
Hence, using these fields, the spectral function is
\begin{align}
\label{eq:A}
A_r(\omega)
&=\int_{-\infty}^\infty dt~e^{i\omega t}\braket{[\tilde{\rho}_r(x=0,t)\tilde{\rho}_r(x=0,0)]}_r \nonumber\\
&=\sum_{n>0}\frac{2\omega_{k_n}}{vL}[\delta(\omega-\omega_{k_n})+\delta(\omega+\omega_{k_n})] 
=\frac{2\omega}{\pi v^2},
\end{align}
where $\omega_{k_n}=vk_n=nv\pi/L$.

\subsection{Fluctuations}

The auto-correlation function of $\tilde{\rho}_r(x=0)=\rho_r$ at the same time is, using the spectral function~\eqref{eq:A},
\begin{align}
\braket{\rho_r^2}
&=S_r(t=0)=\int_{-\infty}^\infty \frac{d\omega}{2\pi}\frac{A_r(\omega)}{1-e^{-\omega/T_r}} \nonumber\\
&=\frac{1}{(\pi v)^2}\int_0^\infty d\omega~\omega[2n_r(\omega)+1].
\end{align}
Therefore, we obtain the fluctuations of $\rho_r$ as
\begin{align}
\label{eq:fluc}
\braket{:\tilde{\rho}_r^2(x=0):}=\frac{T^3}{3v^2}.
\end{align}

By plugging the spectral function~\eqref{eq:A} and the fluctuations~\eqref{eq:fluc} into each contribution, $J_2$, $J_4^{(a)}$, and $J_4^{(b)}$, we obtain the steady-state heat current~\eqref{eq:JIR} in the main text. 

\bibliography{ref}

\end{document}